\documentclass[11pt,twocolumn]{aastex62}

\revised{\today}
\submitjournal{ApJL}

\shorttitle{Sagittarius stream alpha abundances}
\shortauthors{Carlin et al.}

\begin{document}

\title{Chemical abundances of hydrostatic and explosive alpha-elements in Sagittarius stream stars}

\correspondingauthor{Jeffrey L. Carlin}
\email{jcarlin@lsst.org}

\author[0000-0002-3936-9628]{Jeffrey L. Carlin}
\affil{LSST, 950 North Cherry Avenue, Tucson, AZ 85719, USA}

\author{Allyson A. Sheffield}
\affiliation{Department of Natural Science, City University of New York, LaGuardia Community College, Long Island City, NY 11101, USA}

\author{Katia Cunha}
\affiliation{Steward Observatory, 933 North Cherry Avenue, Tucson, AZ 85721, USA; Observat\'{o}rio Nacional/MCTI, Rio de Janeiro, Brazil}

\author{Verne V. Smith}
\affiliation{National Optical Astronomy Observatory, 950 North Cherry Avenue, Tucson, AZ 85719, USA}

\begin{abstract}
We analyze chemical abundances of stars in the Sagittarius (Sgr) tidal stream using high-resolution Gemini+GRACES spectra of 42 members of the highest surface-brightness portions of both the trailing and leading arms. Targets were chosen using a 2MASS+WISE color-color selection, combined with Large Sky Area Multi-Object Fibre Spectroscopic Telescope (LAMOST) radial velocities. 
In this Letter, we analyze [Fe/H] and $\alpha$-elements produced by both hydrostatic (O, Mg) and explosive (Si, Ca, Ti) nucleosynthetic processes.
The average [Fe/H] for our Sgr stream stars is lower than that for stars in the Sgr core, and stars in the trailing and leading arms show systematic differences in [Fe/H]. Both hydrostatic and explosive elements are depleted relative to Milky Way (MW) disk and halo stars, with a larger gap between the MW trend and Sgr stars for the hydrostatic elements. Chemical abundances of Sgr stream stars show similar patterns to those measured in the core of the Sgr dSph. We explore the ratio of hydrostatic to explosive $\alpha$-elements [$\alpha_{\rm h/ex}$] (which we refer to as the ``HEx ratio'').
Our observed HEx ratio trends for Sgr debris are deficient relative to MW stars. Via simple chemical evolution modeling, we show that these HEx ratio patterns are consistent with a Sgr IMF that lacks the most massive stars. 
This study provides a link between the chemical properties in the intact Sgr core and the significant portion of the Sgr system's luminosity that is estimated to currently reside in the streams.
\end{abstract}




\section{Introduction} \label{sec:intro}
The Sagittarius (Sgr) dwarf spheroidal (dSph) galaxy and its sweeping tidal tails provide direct evidence of ongoing mergers in the Milky Way (MW). Sgr is a recently infallen, currently disrupting dwarf spheroidal galaxy, with roughly 70\% of the luminosity of the Sgr system residing in the tidal streams \citep{Niederste-Ostholt2010}. Thus, the Sgr streams are critical for understanding chemical evolution in dwarf galaxy environments and the process of satellite accretion. 
Studies of $\alpha$-elements in metal-rich red giant stars in the Sgr core (\citealt{Bonifacio2004}, \citealt{Monaco2005}, \citealt{Sbordone2007}) point toward a steep (i.e., ``top-light'') initial mass function (IMF). 
\citet{McWilliam2013} analyzed 
abundances of Sgr core RGB stars and found a deficiency in [Mg/Ca] as a function of [Fe/H] for Sgr core stars relative to the MW disk, which suggests a lack of massive stars as nucleosynthetic yields (e.g., Mg) increase with stellar mass; they interpreted their results as evidence for a top-light IMF in Sgr. In the largest study to date of Sgr core red giants, \citet{Hasselquist2017} culled stars from the Apache Point Observatory Galactic Evolution Experiment (APOGEE \citealt{majewski17}) and confirmed the depletion in $\alpha$-elements; using a simple chemical formation model, they show that the results are consistent with a lack of the most massive Type II supernova progenitors.

There are relatively few studies of high-resolution chemical abundances in the Sgr stream.
\citet{Monaco2007a}, \citet{Chou2010a}, and \citet{Keller2010a} analyzed Sgr M-giant stars, and found that stream stars are, on average, more metal-poor than typical Sgr core stars, and are deficient in $\alpha$-elements relative to MW populations at similar metallicities. 

The abundances of $\alpha$-elements in dwarf galaxies are important probes of their star formation timescales. Because $\alpha$-elements are produced in type II supernovae, 
their abundances provide a measure of how efficiently a system produced and retained massive-star products before the onset of type Ia supernovae. In this Letter, we further separate the $\alpha$-elements into those produced internally in massive stars via hydrostatic burning (O and Mg), and those predominantly synthesized in the supernova event (the explosive elements Si, Ca, and Ti). By exploring the ratio of hydrostatic to explosive elements in 42 Sgr stream stars, we show that Sgr must have had a smaller fraction of very massive stars than typical MW stellar populations. 


\section{Data Acquisition and Reduction} \label{sec:data}

\subsection{Target Selection of Sagittarius Stream Stars}

The Sgr streams contain ample populations of M-giant stars \citep[e.g.,][]{Majewski2003}, which are otherwise rare in the Galactic halo where typical metallicities are too low for old RGB stars to reach temperatures cool enough to be classified as M-stars. In this Letter we focus on cool, late-K to early-M-type giants, which are intrinsically bright and thus more amenable to high-resolution follow up. 

\begin{figure}[!t]
\includegraphics[width=1.0\columnwidth]{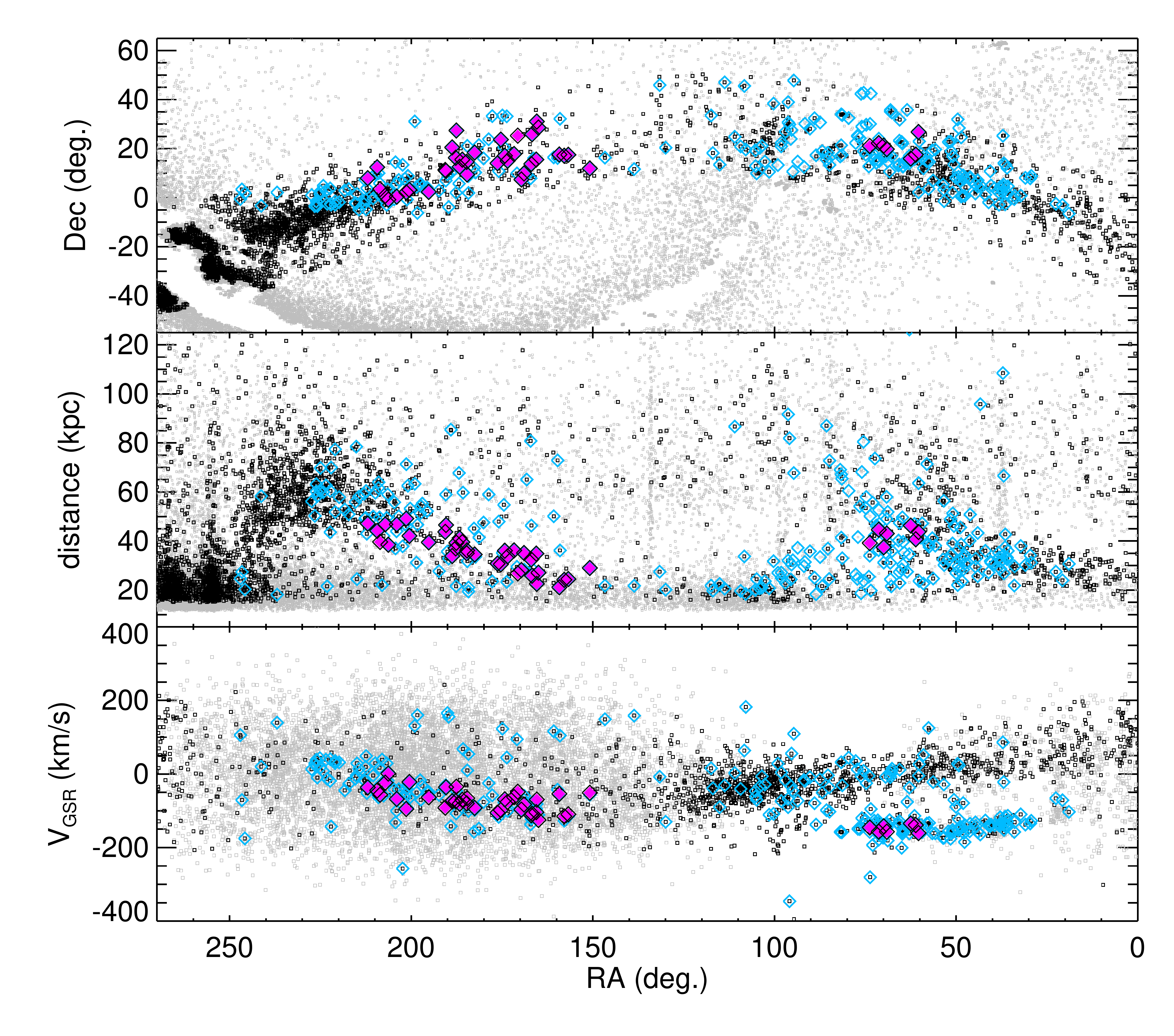}
\caption{Illustration of our method of selecting Sgr stream candidates. Gray points in the top two panels represent an all-sky sample of candidate M-giants selected using the WISE+2MASS color-selection criteria of \citet{Li2016a}.
A sub-selection based on proximity to the Sgr stream plane, distance, color, and proper motion is shown as black points in the upper two panels. Blue points in all panels are M-giant Sgr candidates that matched sources observed in LAMOST DR3. 
The 43 candidates we selected for Gemini+GRACES follow-up are shown as magenta diamonds.}\label{fig:sgrcands2016A}
\end{figure}

Candidates for high-resolution follow up were selected from an all-sky catalog of M-giant candidates using the \citet[][]{Li2016a} 2MASS+WISE photometric selection criteria.
We further require stars to be near the Sgr debris plane (using the Sgr coordinates of \citealt{Majewski2003}), with low extinction and minimal proper motion, and at distances consistent with known properties of the Sgr stream. We match this subset of Sgr M-giant candidates to the Large Sky Area Multi-Object Fibre Spectroscopic Telescope (LAMOST; \citealt{Cui2012,Zhao2012}) DR3 spectroscopic catalog\footnote{\url{http://dr3.lamost.org/}} to obtain velocities, and use known velocity trends in the Sgr stream \citep{Belokurov2014} to choose candidates with velocities consistent with Sgr stream membership, as shown in Figure~\ref{fig:sgrcands2016A}. A subset of 43 velocity-selected candidates constitutes our final sample; all but one of these 43 stars have been confirmed by our analysis to be likely Sgr stream members.

\subsection{Gemini/GRACES Observations}

We obtained spectra of 43 Sagittarius stream K/M-giant candidates
during the 2016A semester via queue-mode observations using Gemini Remote Access to CFHT ESPaDOnS Spectrograph \citep[GRACES ;][]{GRACES_Tollestrup2012,GRACES_Chene2014}.
In the single-fiber mode, GRACES provides spectral resolution $R \equiv \lambda/\Delta\lambda \sim67500$, spanning $4000 \lesssim \lambda \lesssim 10000$~\AA~over 35 echelle orders.
Exposure times ranged from 18-67 minutes, split over 3 exposures per star.
Target magnitudes are between $11.2 < K_{\rm S} < 12.6$ ($15.5 < g_{\rm PS1} < 18.1$ in PanSTARRS). 
We measure signal-to-noise ($S/N$) ratios from the resulting spectra of $\sim40-70$ per pixel (mean: 58.4) between wavelengths 7500-7560~\AA.
We also observed the K-giant Arcturus ($\alpha$~Boo) as a benchmark star.

The data were reduced using the pipeline Data Reduction and Analysis for GRACES\footnote{\url{https://github.com/AndreNicolasChene/DRAGRACES}.} (DRAGRACES; Chen\'e et al. (in prep), v1.1, Zenodo, \url{https://doi.org/10.5281/zenodo.817613}, as developed on github), an IDL program that performs all standard echelle data reduction and extraction tasks (e.g., fitting of aperture traces, scattered-light corrections). 
For each target, the three 1D spectra were shifted and summed using the IRAF task $scombine$.

\section{Derivation of chemical abundances} \label{sec:params}
\subsection{Atomic transition data}

The line list for Fe, O, Mg, Si, Ca, and Ti from \citet{McWilliam2013} was adopted and supplemented with lines from \citet{Friel2010} and \citet{Yong2016}.
The $\log{gf}$ values for each transition (except for the O line, for which we adopt the value from \citealt{Caffau2013}) were placed on a solar scale by measuring their equivalent widths (EWs) and resulting abundances in a high-resolution solar spectrum, then adjusting each $\log{gf}$ value to reproduce the known solar abundances from \citet{Asplund2009}. Because these are cool stars, we avoid regions that could potentially be contaminated by TiO bands.

\subsection{Stellar atmospheric parameters and abundances}
We used SMH \citep{smh}, which is a Python wrapper of the LTE spectral synthesis code MOOG \citep{sneden73}, to continuum normalize and analyze the spectra. 
Radial velocities were measured by cross-correlating a single echelle order with a solar spectrum. 
Stellar atmospheric parameters ($T_{\rm eff}$, microturbulence ($\xi$), [Fe/H], $\log{g}$) were found by iteratively applying standard curve-of-growth analysis to our measured EWs, using synthetic spectra generated with \citet{castelli03} atmospheric models, and interpolating $\log{g}$ values to Dartmouth isochrones \citep{dotter08}.
 
\begin{figure}[!t]
\includegraphics[width=1.0\columnwidth,trim=0.25in 0.25in 0.5in 0.0in]{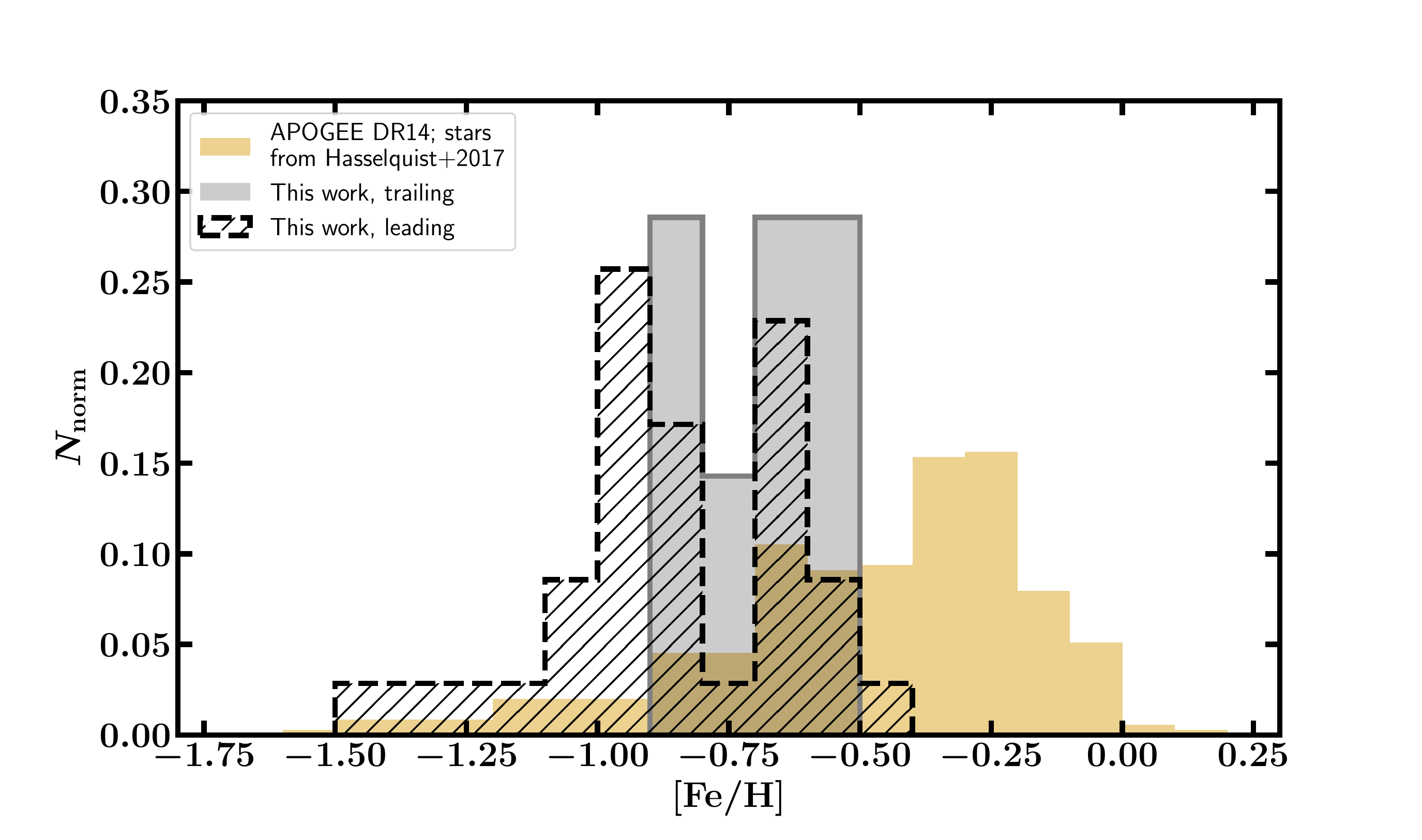}
\caption{Metallicities ([Fe/H]) for Sgr stars. Stars in the trailing stream are the filled grey region; those in the leading are the hashed histogram. Stars in the Sgr core are shown in gold and are APOGEE DR14 results for stars from \citet{Hasselquist2017}.}\label{fig:sgr_feh}
\end{figure}

Once the stellar atmospheric parameters were determined, we ran the MOOG $abfind$ driver to determine the abundances of Fe, O, Mg, Si, Ca, and Ti. More details of the analysis will be presented in a forthcoming work.
We derived metallicities ([Fe/H]) for each star as the mean from 50 Fe~I lines.
The median uncertainty on [Fe/H], based on the scatter of individual line measurements, is 0.09 dex. Typical uncertainties on temperature and surface gravity are $\sigma_{T_{\rm eff}} \sim 115$~K, $\sigma_{\log{g}} \sim 0.18$~dex.

\subsection{Metallicities of Sgr stream stars}

The distribution of measured [Fe/H] for Sgr stream stars is shown in Figure~\ref{fig:sgr_feh}. 
The seven stars in the trailing tail have median $\langle$[Fe/H]$\rangle = -0.68$ with scatter of 0.11 dex, while for the 35 leading arm stars these values are $\langle$[Fe/H]$\rangle = -0.89$, $\sigma_{\rm [Fe/H]} = 0.24$~dex. A two-sided K-S test comparing the two distributions yields a $p$-value of 0.01, rejecting the null hypothesis that they are drawn from the same population.
This systematic difference 
likely reflects differences in the time that these stars became unbound from the Sgr dSph, with tidal dissolution proceeding from the least tightly bound stars to those nearer the core. 
The results of $N$-body modeling of Sgr tidal disruption by \citet{LM10} suggest that trailing tail stars in our sample were stripped within the past $\sim0.7-3$~Gyr, while the leading arm stars that we observed should be almost exclusively from earlier debris stripped $\sim2.7-5.0$~Gyr ago. The average difference in metallicity as a function of debris age likely reflects metallicity gradients that were originally present in the Sgr core (as suggested by, e.g., \citealt{Bellazzini2006a,Chou2007, Keller2010a}). A population gradient in/near the Sgr core has been measured photometrically \citep{Bellazzini1999,Alard2001,McDonald2013}, and an [Fe/H] gradient seen spectroscopically \citep{Majewski2013,Hasselquist2017}. 

Our average M-giant metallicity in both the leading and trailing streams is markedly lower than the peak of [Fe/H]$\sim -0.3$ from M-giants in the Sgr core \citep[e.g., ][]{Smecker-Hane2002,Monaco2005,Hasselquist2017}. The most metal-rich stellar populations present in the Sgr core, which formed $\sim0.7$~Gyr ago \citep{siegel07}, are not found in the portions of the streams that we have observed. This suggests that at minimum the debris that we are studying were stripped from the Sgr progenitor more than 0.7~Gyr ago.


\section{Alpha-element Abundances in the Sagittarius Stream} \label{sec:alphas}

It is well established that the most metal-rich stellar populations in dwarf galaxies are deficient in $\alpha$-elements relative to stars of the same metallicity in the Milky Way (e.g., \citealt{Tolstoy2009,Kirby2011} and references therein). This is typically shown by combining the measured abundances of the most readily available $\alpha$-element species to examine an average $[\alpha$/Fe] vs. [Fe/H]. 
Early chemical evolution in a system is dominated by Type II supernovae (SNeII) 
while at later times core-collapse SNeIa begin to dominate, diluting $[\alpha$/Fe] by contributing a larger fraction of Fe. Thus the downturn in [$\alpha$/Fe] vs. [Fe/H]
is a sign of the enrichment level reached by a system at early times.

\begin{figure}[!t]
\includegraphics[width=1.0\columnwidth, trim=0.2in 0.25in 0.6in 0.0in, clip]{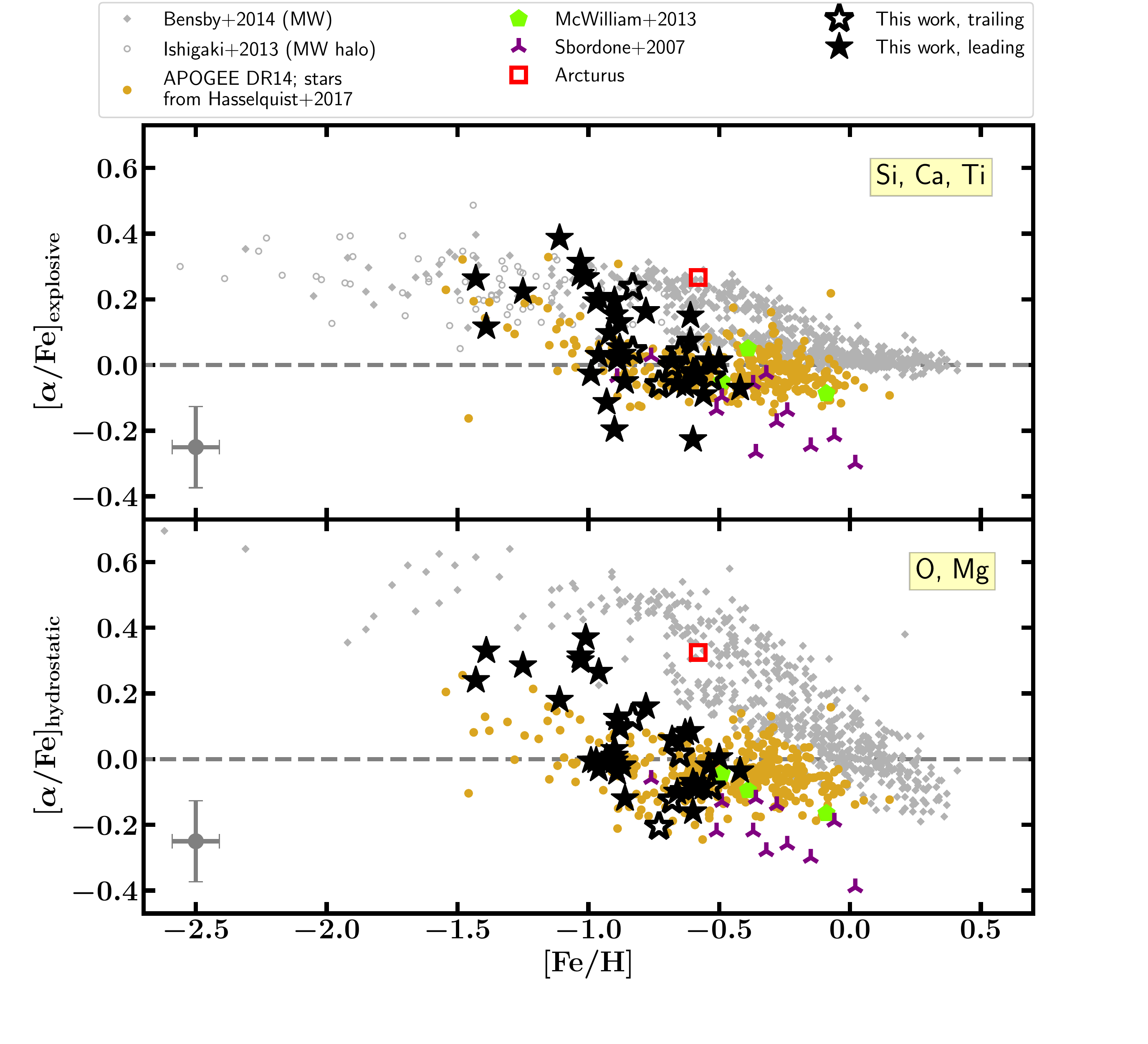}
\caption{Measured explosive and hydrostatic $[\alpha$/Fe] ratios for Sgr stream stars (upper and lower panels, respectively), compared to the Galactic thin/thick disks \citep{Bensby2014}, thick disk/halo \citep{Ishigaki2012,Ishigaki2013}, and Sgr dSph stars \citep{Sbordone2007,McWilliam2013,Hasselquist2017}. The disk K-giant Arcturus is the red square. 
In both panels, Sgr stream stars' (large black stars) abundances look similar to those in the present-day Sgr dSph. }\label{fig:sgr_alpha}
\end{figure}

\begin{figure*}[!t]
\includegraphics[width=1.0\textwidth, trim=0.25in 0.25in 0.75in 0.25in]{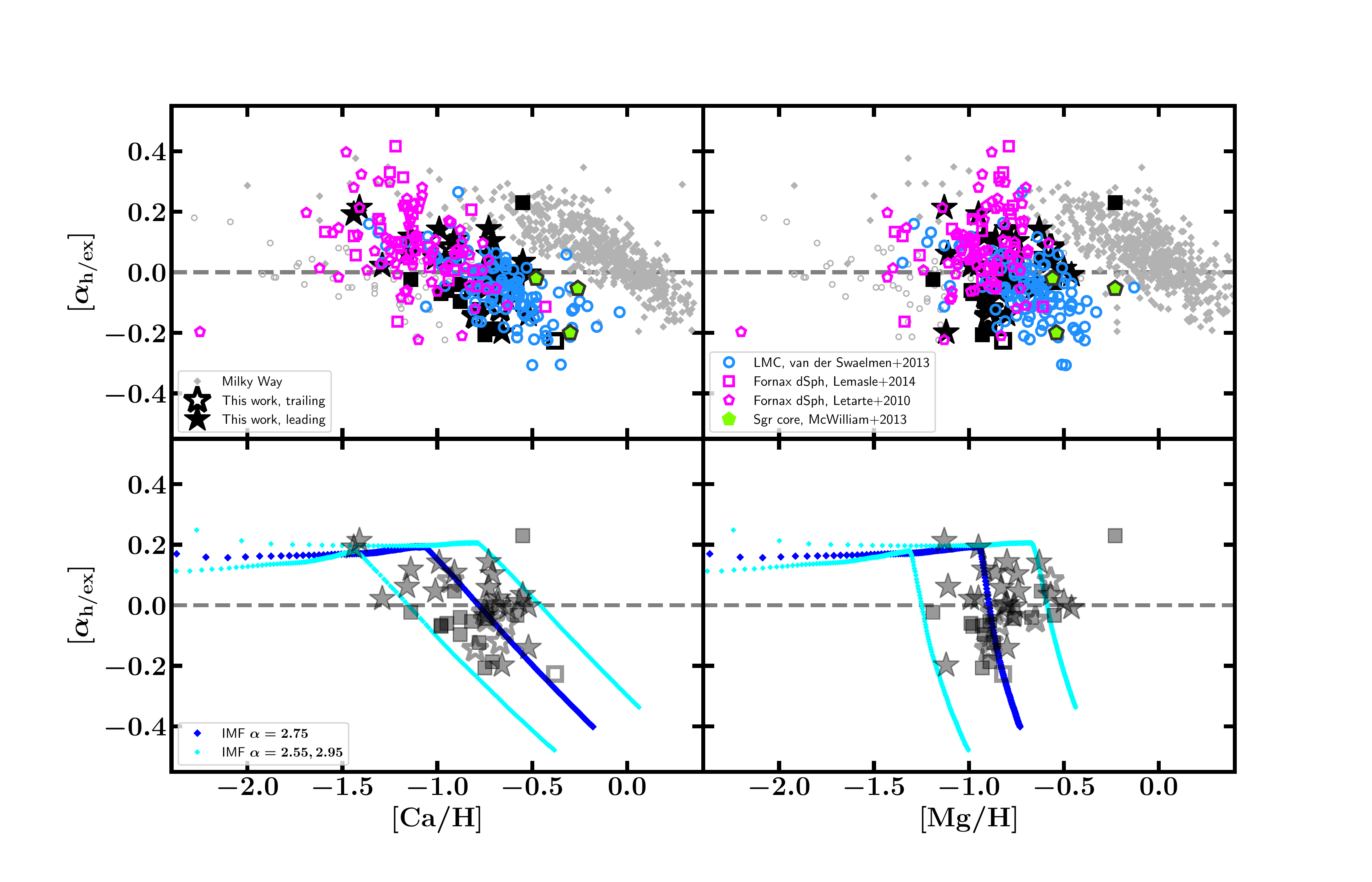}
\caption{The ``HEx ratio,'' [$\alpha_{\rm h/ex}$]: the ratio of the average of hydrostatic $\alpha$-elements Mg and O to the average of explosive $\alpha$-elements Si, Ca, and Ti. Left panels show this ratio as a function of [Ca/H], while the right panels depict [$\alpha_{\rm h/ex}$] vs. [Mg/H]. Our Sgr stream targets are large black/gray stars in each panel; stars with no O measurement include only Mg in the ratio, and are shown as squares. The upper row compares to MW stars from \citet{Bensby2014} and \citet[gray points]{Ishigaki2012,Ishigaki2013}, 
to LMC abundances \citep[blue open circles]{VanderSwaelmen2013}, and to stars in the Fornax dSph \citep[magenta points; which include only Mg in $\alpha_{\rm hyd}$]{Letarte2010,Lemasle2014} and 3 stars in the Sgr core \citep{McWilliam2013}. Sgr stream [$\alpha_{\rm h/ex}$] abundance patterns closely follow those of the LMC and Fornax, and are well separated from MW populations. Lower panels show only Sgr stream stars, with the results of \textit{flexCE} \citep{flexce} modeling of a stellar population with strong outflows, IMF power-law slope $\alpha = 2.75$, and SNIa time delay of 1.2~Gyr overlaid as blue dots; cyan points show the effect of changing the IMF slope by $\pm0.2$. }\label{fig:sims_hex}
\end{figure*}

\subsection{Not all $\alpha$-elements are the same: hydrostatic vs. explosive elements}

While the $\alpha$-elements O, Mg, Si, Ca, and Ti are often treated as a homogeneous group, they do not all originate from the same nucleosynthetic process. Thus it is more informative to separate O and Mg, produced mainly via hydrostatic burning of C and Ne in massive stars, from Si, Ca, and Ti, which are primarily synthesized in the SNII explosion. 

Figure~\ref{fig:sgr_alpha} shows the average abundance of the explosive elements Si, Ca, and Ti ([$\alpha/{\rm Fe}]_{\rm explosive}$; upper panel), and the average of hydrostatic elements O and Mg ([$\alpha/{\rm Fe}]_{\rm hydrostatic}$; lower plot) for our 42 Sgr stream stars. For comparison we include a local Galactic disk sample from \citet{Bensby2014} thick-disk/halo stars from \citet{Ishigaki2012,Ishigaki2013}, 
and abundances of Sgr core stars from the literature (\citealt{Sbordone2007}, \citealt{McWilliam2013}, and \citealt{Hasselquist2017}). Our measured abundances for Sgr stream stars in both panels mainly follow the patterns of stars in the core, although 
most of the stars in these previous studies are more metal rich than our sample.
The trends in [$\alpha/{\rm Fe}]_{\rm explosive}$ and [$\alpha/{\rm Fe}]_{\rm hydrostatic}$ vs. [Fe/H] clearly extend from the higher-metallicity Sgr core stars to our lower-[Fe/H] stream members.
More metal-poor stream stars would be expected to come from, on average, older stellar populations (corresponding to the metal-poor population and the oldest of the intermediate-age bursts as discussed in \citealt{siegel07}, for example). 
Although models \citep{LM10} suggest that debris in our study were stripped from Sgr $\sim1-5$~Gyr ago, our results show that we can clearly relate them to the populations still present in the core via their chemical abundance trends.

The gap between Sgr stars and the MW disk trends is larger in [$\alpha/{\rm Fe}]_{\rm hydrostatic}$ than it is in [$\alpha/{\rm Fe}]_{\rm explosive}$. 
\citet{McWilliam2013} and \citet{Vincenzo2015} similarly noted that together with, e.g., the Eu ($r$-process) abundances, the relative amounts of hydrostatic/explosive elements argue for a lack of the most massive stars in the Sgr system relative to the MW. 

\section{The ``HEx ratio'' and a simple chemical evolution model for the Sgr stream}

To illustrate the points made above about the different origins of hydrostatic and explosive alpha-elements, we take the ratio of $\alpha_{\rm hydrostatic} =$~([Mg/Fe]+[O/Fe])/2 (or $\alpha_{\rm hyd} =$~[Mg/Fe] for the 15 stars that do not have O measurements)
to $\alpha_{\rm explosive} =$~([Si/Fe]+[Ca/Fe]+[Ti/Fe])/3.0, which we formulate as [$\alpha_{\rm h/ex}] = \alpha_{\rm hydrostatic} - \alpha_{\rm explosive}$ (and refer to as the ``HEx ratio''). We show [$\alpha_{\rm h/ex}$] vs. [Ca/H] and [Mg/H] for our Sgr stream stars in Figure~\ref{fig:sims_hex}. Note that 
the HEx ratio focuses solely on $\alpha$-elements, independent of Fe.
Sgr stream stars show a clear pattern in Figure~\ref{fig:sims_hex}, with a higher abundance of hydrostatic $\alpha$-elements at lower [Ca/H], which decreases toward higher [Ca/H] such that most of the Sgr stream stars have [$\alpha_{\rm h/ex}] < 0$, or a higher abundance of explosive relative to hydrostatic elements. The HEx ratio decreases in the [Mg/H] panels, but at a nearly constant value of [Mg/H], suggesting that little enrichment in Mg is happening after the downturn.

The upper panels of Figure~\ref{fig:sims_hex} compare our measured Sgr stream HEx ratios to those of Milky Way stars, LMC stars, RGB members of the Fornax dSph, and 3 stars from the Sgr core (the only available Sgr core stars with suitable abundances to calculate [$\alpha_{\rm h/ex}$]). The trend followed by Sgr stream stars 
is much different from that of MW stellar populations. The lowest [Ca/H] Galactic halo populations have roughly Solar [$\alpha_{\rm h/ex}$] ratios, which rise slightly with increasing [Ca/H], transitioning into a declining trend with [Ca/H] at thick- and thin-disk metallicities. Sgr, on the other hand, starts above the MW halo in the HEx ratio at low [Ca/H], and monotonically declines (on average) to values well below the disk at higher [Ca/H]. Furthermore, Sgr stream stars look similar to both the Fornax dSph (at the metal-poor end) and the LMC. Indeed, the works to which we compare for the LMC \citep{VanderSwaelmen2013} and Fornax \citep{Letarte2010,Lemasle2014} all conclude that there is chemical evidence for a paucity of massive stars (i.e., a top-light IMF) in these systems.

In the left panels of Figure~\ref{fig:sims_hex}, Ca continues rising, while Mg, in contrast, remains roughly constant among the M-giant sample that we probe. This fits with the overall picture we are discussing, because Ca is produced in all SNII events (in the explosion itself), whereas Mg is primarily synthesized hydrostatically only in the interiors of stars that are massive enough to ignite burning of C and Ne. Finally, we note that $\sim20\%$ of Ca is also produced in SNeIa \citep{Iwamoto1999}, which contribute little Mg; this partly accounts for the difference in slope above [Ca/H] or [Mg/H]$\gtrsim-1.0$.

We examine the plausibility of the top-light IMF scenario using the \textit{flexCE} chemical evolution model \citep{flexce}.\footnote{\url{https://github.com/bretthandrews/flexCE}} 
The \textit{flexCE} model that best reproduces the Sgr stream stars in Figure~\ref{fig:sims_hex} is a 2~Gyr star formation episode, including stars from 0.1-100~$M_\odot$ with a power-law IMF of slope $\alpha = 2.75$ (where $dN/dm \propto m^{-\alpha}$; $N$ is the number of stars of mass $m$; $\alpha = 2.35$ represents a ``standard'' Salpeter IMF). The onset of SNeIa was delayed until 1.2 Gyr, with no inflows, a substantial wind outflow ($\eta_{\rm wind} = 9$; as required in the Sgr chemical evolution models of \citealt{Lanfranchi2006}), and a standard Kennicutt-Schmidt star formation law with power-law slope of 1.4 \citep{Kennicutt1998} and 10~Gyr gas depletion timescale. 

The results of this model are plotted with our observed Sgr abundances in the lower panels of Figure~\ref{fig:sims_hex}. Blue points show the model described above, with IMF slope $\alpha = 2.75$, and cyan points represent the same model, but with IMF slope varied by $\pm0.2$. The model reproduces the observed trends well, with the ``knee'' at [Ca/H]$\sim -1.0$ corresponding to the onset of type Ia SNe. 
The abrupt drop in the hydrostatic/explosive ratio vs. [Mg/H] likely corresponds to a decrease in the number of the most massive stars,
which are required to burn C and Ne in order to form the hydrostatic element Mg (note that O requires even higher temperatures, and, if available for all stars, would be an even more sensitive probe of the relative numbers of very massive stars). After this point, the abundance of [Mg/H] remains roughly constant, while the lower-mass Type II SNe progenitors continue to synthesize explosive $\alpha$-elements during their abrupt demise as SNeII, thus lowering [$\alpha_{\rm h/ex}$].

\section{Conclusions} \label{sec:discussion}

We present results from the largest sample of high-resolution, multi-element chemical abundances yet obtained in the Sagittarius stream.  
Based on $\alpha$-element abundances of 42 Sgr stream stars, we conclude the following.

\begin{itemize}
  \item Stars in the Sgr stream(s) have $\alpha$-abundance patterns that look similar to those of stars in the Sgr dSph core.
  \item Leading arm stars in our sample are on average more metal poor than those in the trailing tail. Stars in both regions follow similar overall $\alpha$-abundance trends to those in the Sgr core. 
  \item Sgr stream stars are deficient (relative to Galactic stellar populations at similar [Fe/H]) in $\alpha$-elements. Furthermore, the deficiency is more pronounced for species that are formed in hydrostatic processes than for those originating in explosive synthesis. This suggests that the Sgr dSph lacked the most massive SNII progenitors, which would be the sites of hydrostatic synthesis of elements such as O and Mg, confirming the suggestion by \citet{McWilliam2013} and \citet{Vincenzo2015} that the Sgr dwarf had a top-light IMF.
  \item The $\alpha$-element abundance patterns we see for the Sgr stream (and core) are similar to those observed in the LMC and the MW dSph Fornax. This similarity to the LMC suggests that Sgr, in spite of its current luminosity similar to that of Fornax, was once much more massive than it currently is. The fact that Sgr was more massive in the past is not surprising, given the Sgr tidal streams that stretch over a huge swath of sky, but our results 
suggest that Sgr was once as massive as the LMC (as also suggested based on chemical similarities by \citealt{Mucciarelli2017}, and by \citealt{Gibbons2017} based on velocity dispersions). This, in turn, would have ramifications for modeling the orbital history of Sgr in the MW system. 
  \item We model the chemical evolution of Sgr in order to reproduce the abundances of the old M-giants in the stream, and find that their $\alpha$-abundance patterns are consistent with a brief, ancient star formation episode, with strong outflows, a top-light IMF, and a long time delay before SNIa onset. 

\end{itemize}

In combination with detailed chemical abundances of Sgr core stars, our analysis provides the framework for understanding the chemical evolution of the Sgr dwarf galaxy, and confirms the plausibility of identifying Sgr tidal debris via chemical signatures that readily distinguish Sgr stars from MW stellar populations.

\acknowledgments

We thank Kathryn Johnston for hosting a collaboration meeting at Columbia University in support of this project, Andr\'e-Nicolas Chen\'e for assistance with DRAGRACES, and Jing Li for sharing her catalog of 2MASS+WISE M-giants. We are also grateful for the hospitality and stimulating discussions at the Center for Computational Astrophysics, Flatiron Institute, New York, NY. We thank the referee for a careful review.

This work is based on observations 
obtained with ESPaDOnS, located at the Canada-France-Hawaii Telescope (CFHT). 
The Guoshoujing Telescope (LAMOST) is a National Major Scientific Project built by the Chinese Academy of Sciences. Funding for the project has been provided by the National Development and Reform Commission. LAMOST is operated and managed by the National Astronomical Observatories, Chinese Academy of Sciences.

\end{document}